\documentclass[grl, draft]{agu2001} 

\usepackage{lineno}

\usepackage{graphicx, latexsym, amssymb, color}

\def\lsim{~\rlap{$<$}{\lower 0.5ex\hbox{$\sim$}}}

\authorrunninghead{JUANICO et al.}

\titlerunninghead{AVALANCHE STATISTICS OF DRIVEN GRANULAR SLIDES...}

\authoraddr{cmonterola@nip.upd.edu.ph,cmonterola@gmail.com}

\linenumbers*[1]

\begin{document}


%


%


%



\setkeys{Gin}{draft=false}

%




%


%


%


\title{Avalanche Statistics of Driven Granular Slides in a Miniature Mound}

%




%


%





\author{D.E. Juanico, A. Longjas, R. Batac, and C. Monterola}

\affil{National Institute of Physics, University of the Philippines
, Quezon City, Philippines}

\begin{abstract}

We examine avalanche statistics of rain- and vibration-driven
granular slides in miniature sand mounds. A crossover from power-law
to non power-law avalanche-size statistics is demonstrated as a
generic driving rate $\nu$ is increased. For slowly-driven mounds,
the tail of the avalanche-size distribution is a power-law with
exponent $-1.97\pm 0.31$, reasonably close to the value previously
reported for landslide volumes. The interevent occurrence times are
also analyzed for slowly-driven mounds; its distribution exhibits a
power-law with exponent $-2.670\pm 0.001$.

\end{abstract}


%


%





%


\begin{article}


%


%



\emph{Introduction}.---Landslides are the movement of a mass of
rock, debris, or earth down a slope triggered by a variety of
natural factors, ranging from rainfall to volcanic activity. On 17
February 2006, a series of mudslides caused widespread damage and
loss of life in Southern Leyte, Philippines. The deadly landslides
followed a ten-day period of persistent downpour and a minor
($M2.6$) earthquake~\citep{Catane2007}.


One practical approach in analyzing the underlying physical
processes that generate landslide statistics is numerical modeling.
One class of models, based on \emph{self-organized criticality}
(SOC) \citep{HergartenNeugebauer1998, Piegari2006}, hint at some
possible mechanisms yielding the observed statistics. SOC is a
theory underlying the spontaneous emergence of critical-like
behavior (i.e., power laws and critical exponents) in systems for
which the timescales between buildup and release of stress are
separated, and for which the stress-transfer mechanism is generally
nonconservative~\citep{Juanico2007a,Juanico2007b,JuanicoMonterola2007}.
SOC concepts have aroused great interest in the study of granular
matter~\citep{Jaeger1989}, a well-known example of which is the
ricepile experiment~\citep{Frette1996}.


The present work examines avalanche statistics of rain- and
vibration-driven granular slides in miniature sand mounds. A
previous study~\citep{KatzAharonov2006} explored slope-failure types
due to horizontal and vertical vibrations in a miniature sandbox. It
was shown that vertical shaking leads to a power-law distribution of
slide-block surface area, although the experiment did not
demonstrate a `rollover' observed in substantially complete,
empirical landslide
inventories~\citep{GuzzettiMalamudTurcotteReichenbach2002,MalamudTurcotteGuzzettiReichenbach2004a,MalamudTurcotteGuzzettiReichenbach2004b}.
In the present study, we incorporate rainfall as a triggering
mechanism and show that its presence allows our model to capture the
rollover. Likewise, we demonstrate experimentally the existence of a
crossover from power-law to non power-law statistics as
theoretically predicted by~\cite{Piegari2006}.

It appears from the data reported
by~\cite{MalamudTurcotteGuzzettiReichenbach2004a} that any type of
trigger (earthquake, rainfall, or snowmelt) yields roughly the same
trend in the landslide-size distribution. Combining any two trigger
types should reasonably yield the same trend in the distribution,
and this is precisely one of the aspects tested in this study. In
addition, the computational model proposed here assumes a generic
triggering mechanism. Thus, rainfall and vibrations have been
introduced to act as concurrent landslide triggers.

\begin{figure}
\noindent\includegraphics[width=20pc]{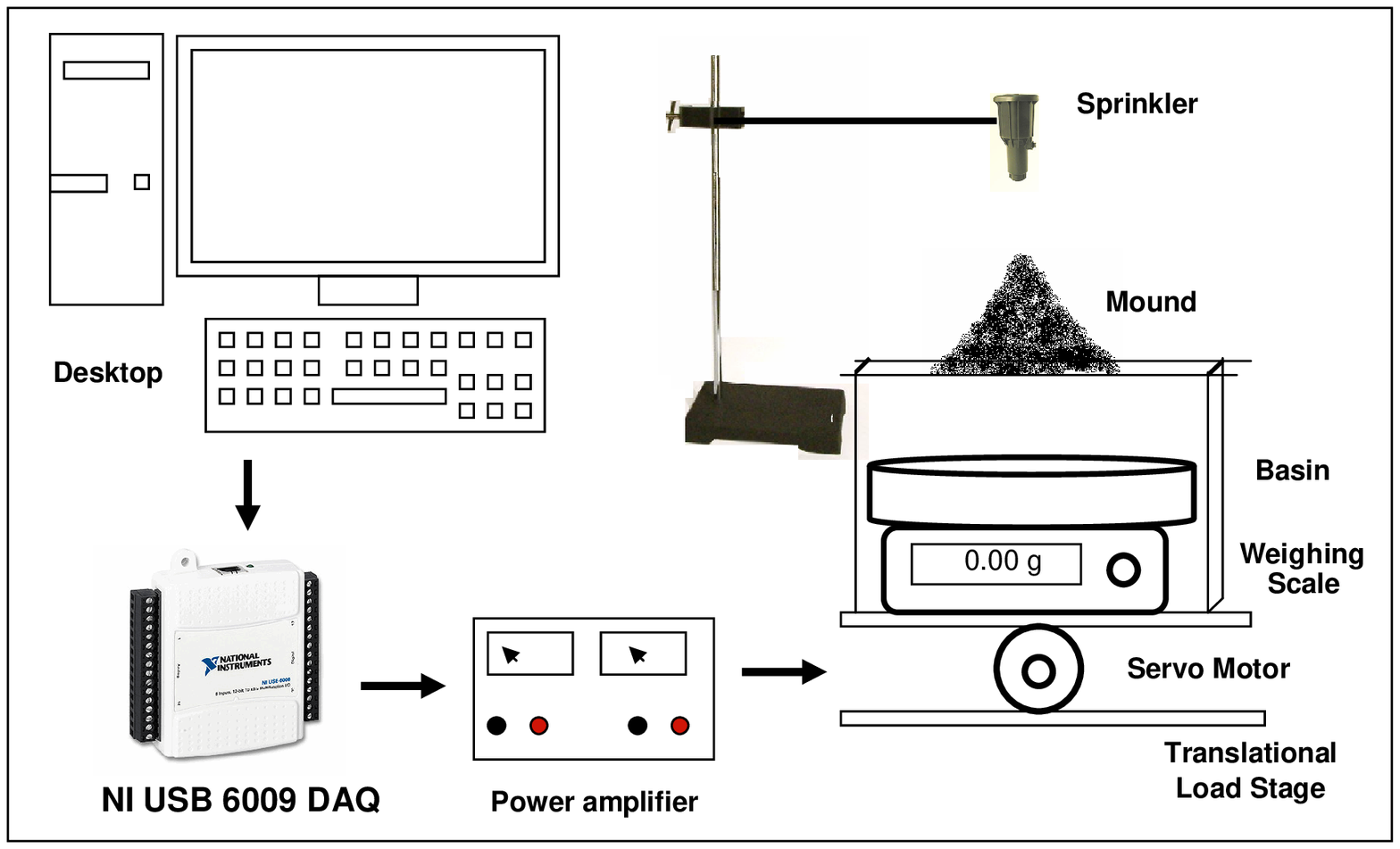}

\caption{Diagram of experimental setup. Sand mound consists of river
sand. Rainfall is simulated by a sprinkler right above the mound
apex. An earthquake is simulated by a translational load stage
imparting horizontal vibrations to the platform where the mound is
placed. Rainfall and vibration are parametrized by pour rate and
vibration frequency, respectively. Different parameter combinations
are listed on Table~\ref{tab:1}. Avalanche size is the mass $M$ of
wet sand falling onto the basin within $20$-ms intervals. $M$ is
measured by a weighing scale at the bottom of the basin, and is
connected to the PC that records measurements in time.}\label{fig:1}
\end{figure}

\emph{Experimental Method.}---The experimental setup
(Figure~\ref{fig:1}) consists of a sand mound (total mass, $1500$~g)
disturbed concurrently by rainfall and vibration. River sand with
irregularly-shaped grains (mean grain mass, $8.8\times 10^{-4}$~g;
mean grain volume, $5.8\times 10^{-4}\,\,\mathrm{cm}^3$) was used.
Initially, the mound is dry and is a near-perfect cone in shape with
a base diameter of $23$~cm and height of $8$~cm (slope angle
$\approx 35^{\circ}$ with respect to horizontal). Water ($200$~ml)
is dispensed quite uniformly over the surface and allowed time to
seep into the mound. The wet mound thus consists of $12$\% water by
weight at the start of the observations.

Rainfall is simulated using a makeshift sprinkler (water capacity,
$75$ ml) placed directly above the mound apex. Water pours out
through a circular orifice of diameters: $4.5$, $10.0$, and
$47.0$~($\pm 0.5$)~mm. Pour rate is the mass of water flowing out
per unit time, and the orifice diameter controls its value. Pour
rate is constant at the following values:
$3.21$~g~$\mathrm{s}^{-1}$; $6.60$~g~$\mathrm{s}^{-1}$; and
$21.09$~g~$\mathrm{s}^{-1}$ over a time interval of $20$, $10$, and
$3$ s, respectively. These time intervals are commensurate with the
length of time the 75-ml water in the sprinkler is depleted.

Horizontal vibrations are applied using a tabletop earthquake
simulator. Horizontal shaking at low to moderate accelerations
generates grain flows of the topmost layer of the slope resulting in
rapid failure-plane development~\citep{KatzAharonov2006}. As shown
in Figure~\ref{fig:1}, a translational-load platform, powered by a
servo motor driven by a USB $6009$ DAQ driver (National
Instruments\texttrademark), imparts the horizontal vibrations having
a sawtooth wave profile. The wave profile is fed into the DAQ driver
via LabVIEW\texttrademark computer interface. The wave is
characterized by a maximum amplitude of $1.5$ cm (with respect to
center) and by frequencies of: $1.8$ Hz, $10$ Hz, and $89$ Hz. Due
to the 3D shape of the mound, slope-parallel and slope-normal
accelerations are both present during shaking.

\emph{Computational Model}.---The underlying physics of the
avalanche statistics of driven granular slides is investigated by
performing numerical experiments of a landslide model proposed
by~\cite{Piegari2006}, defined as follows. A mountain `slope' is
represented as a 2D inclined plane partitioned into a grid
($500\,\,\mathrm{cells}\, \times\, 500\,\, \mathrm{cells}$). Each
cell $k$ is described by a stress parameter $\theta_{k}$ initialized
randomly between $0$ and $1$ from an arbitrarily chosen rectangular
distribution (although the distribution used for initial
randomization does not affect the long-term behavior of the
model~\citep{Piegari2006}). The randomization captures the expected
heterogeneity of stress values in actual mountain slopes. Stress in
the slope builds up over time by means of a localized (i.e., cell
scale) driving: $\theta_{k}\left(t+ \Delta t\right) = \theta_{k}(t)
+ \nu\Delta t$, where $\nu$ is the generic driving rate. For
simplicity, it is assumed that $\nu$ has the same value for all
cells. When a cell $k$ has $\theta_{k}> 1$, it relaxes by
transferring stress to its four nearest neighbors $nn = \left\{up,
down, right, left\right\}$ at different proportions $g_{nn}$. By
virtue of gravity, stress transfer is biased downwards, such that:
$g_{down} > g_{up}$, subject to the constraint
$g_{down}+g_{up}=0.5$; and $g_{left}=g_{right}=0.25$. In this study,
it is assumed that transfer is conservative, so that
$\sum_{nn}g_{nn}=1$. Stress-transfer proceeds until the entire grid
relaxes so that for all $k$, $\theta_{k} < 1$. All consecutive
stress-transfers comprise a landslide at time $t$, and the total
number of collapsing cells at time $t$ is the landslide area $A(t)$.
For correspondence with experiments, although we recognize its
limitations, we assume a landslide mass-area relation: $M \sim
A^{3/2}$, which can be derived by means of dimensional analysis
(i.e., assuming that $M$ is proportional to volume $V$, and then
considering that $V \sim A^{3/2}$~\citep{Hovius1997}).

The novelty of our computational modeling approach is in considering
the gradual flattening of the slope after several landslides have
occurred (as seen in our experiments). An update rule is introduced
to decrease $g_{down}$, as follows: $g_{down}(t+\Delta t) =
g_{down}(t) - 10^{-5}A(t)$, where $A(t)$ is the area of landslide at
time $t$. The value of $g_{up}$ is updated accordingly via the
constraint $g_{down}+g_{up}=0.5$. This modification incorporates the
dynamics of slope evolution which was not realized
by~\cite{Piegari2006}. The generic driving rate $\nu$ defined in the
model corresponds to the experimental parameters, as shown in
Table~\ref{tab:1}.

\begin{table}[h]
\caption{Generic driving rate $\nu$ in terms of experimental
parameters} \label{tab:1}
\begin{tabular}{l||r|r|r} 
  \hline
  Description & Driving rate (no units) $\nu$ & Pour rate (g $\mathrm{s}^{-1}$) & Vibration frequency (Hz) \\\hline
  Slow & $7.50\times 10^{-5}$ & 3.21 & 1.8 \\
  Moderate & $1.25\times 10^{-4}$ & 6.60 & 1.8, 10, 89 \\
  Fast & $7.50\times 10^{-4}$ & 21.09 & 1.8, 10, 89 \\
  \hline
\end{tabular} 
\end{table}

\emph{Results and Discussion}.---Avalanche size is interpreted as
the total mass $M$ of wet sand falling from the mound onto the basin
within a $20$-ms period (temporal resolution of our data-capturing
device). $M$ is measured by a weighing scale (resolution, $0.1$~g)
beneath the basin, as illustrated in Figure~\ref{fig:1}. $M$ is
sometimes contributed to by several distinct `sub-avalanches.' The
tradeoff in this interpretation is that although it resolves mostly
individual avalanches, those that last $>20$~ms may be recorded as
partial sub-avalanches, and if there are more than one avalanche
within any $20$~-ms interval, these will be recorded as one value.

The total observation period is determined by the pour rate and is
based on how long before the entire mound washes out. A total of
$20$, $10$, and $3$ trials were made for slow, moderate, and fast
pour rates, respectively. The fact that sand tends to stick together
when wet (i.e., negative pore pressure)~\citep{KatzAharonov2006}
justifies our assumption that by measuring the mass $M$ of falling
wet sand, the avalanche volume is effectively measured (i.e., $M
\propto V$).
In~\citep{MalamudTurcotteGuzzettiReichenbach2004a,MalamudTurcotteGuzzettiReichenbach2004b},
the distribution of landslide volumes has been deduced from scaling
arguments due to the difficulty in obtaining direct information
about landslide volume from field
data~\citep{MalamudTurcotteGuzzettiReichenbach2004a}. In our
experiment, landslide volume distribution is determined by using
mass as a proxy for volume.

\begin{figure}
\noindent\includegraphics[width=20pc]{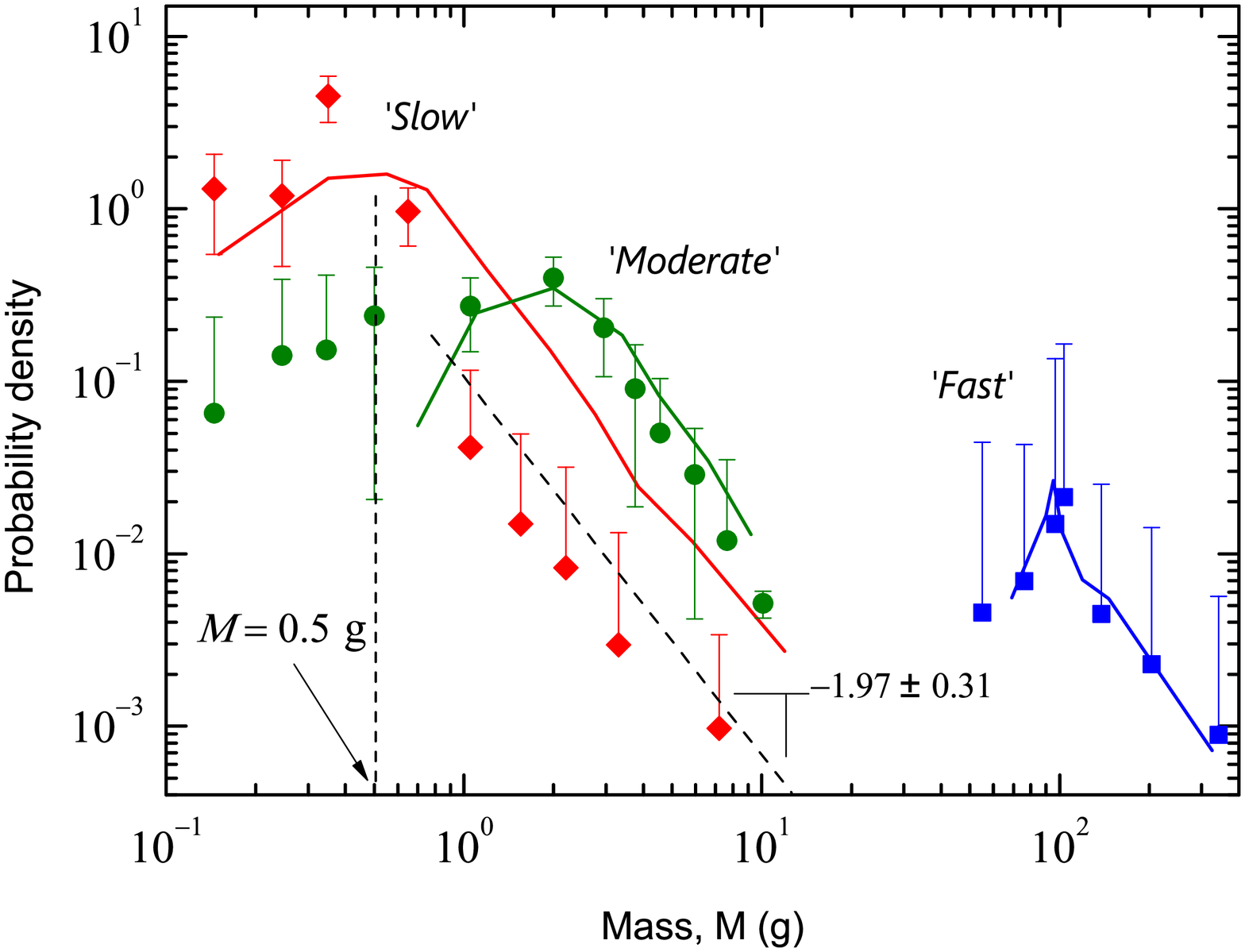}

\caption{(Color online). Probability densities (pdf) of avalanche
sizes. Pdfs for slow (\textcolor{red}{\scriptsize $\blacklozenge$}),
moderate (\textcolor[rgb]{0,0.5,0}{$\bullet$}), and fast
(\textcolor{blue}{\tiny $\blacksquare$}) driving; where the error
bars are $\pm 2\sqrt{n}/\delta$ (equivalent to $\pm 2\sigma$, 95\%
confidence; $n=$ number of values in a bin; and $\delta$ is the bin
width). Corresponding pdfs of avalanche size from numerical
simulations (curves) are overlaid, after converting area into mass
using the scaling relation: $M = (1.0\times 10^{-5}\mathrm{g}\cdot
\mathrm{cell}^{-3/2}) A^{3/2}$. Linear regime ($1.0 < M < 8.0$~g) of
the pdf for a slowly-driven mound fits a power-law with exponent
$-1.97\pm 0.31$. For moderately-driven mounds, the rollover region
of the pdf becomes more prominent as the peak shifts to the right.
For highly-driven mounds, the pdf resembles a normal distribution
centered at a large $M$.}\label{fig:2}
\end{figure}

Figure~\ref{fig:2} illustrates probability densities (pdf) of
avalanche sizes resulting from slow (\textcolor{red}{\scriptsize
$\blacklozenge$}), moderate (\textcolor[rgb]{0,0.5,0}{$\bullet$}),
and fast (\textcolor{blue}{\tiny $\blacksquare$}) driving (as
defined in Table~\ref{tab:1}). Also shown are corresponding pdfs
from numerical simulations of the computational model. The pdfs for
all cases exhibit peaks that shift towards the right as the driving
rate is increased. In particular, The tail of the pdf for a
slowly-driven mound (\textcolor{red}{\scriptsize $\blacklozenge$},
Fig.~\ref{fig:2}) is a power-law with exponent $-1.97$($\pm 0.31$;
standard error of least-squares fit, $df=4$, reduced
$\chi^2=1.14\times 10^{-5}$), determined by curve-fitting the linear
portion (in double log scale) for which $1.0\,\,\mathrm{g}< M <
8.0\,\,\mathrm{g}$. The exponent value is not significantly
different (two-tailed $t$-test: $t = 1.13$, $df=4$, $p=0.32$) from
the value $-1.93$ reported for probability densities of landslide
volumes~\citep{MalamudTurcotteGuzzettiReichenbach2004a}.


A crossover from power-law to non power-law statistics is also
evident in our data. This crossover has been previously recognized
by~\cite{Piegari2006} as an effect of increasing the generic driving
rate $\nu$. We attribute the crossover to timescale separation.
Based on the model, the timescale separation is the ratio between
the timescale $\nu^{-1}$ of stress changes on any site due to
driving and the timescale of the longest avalanche which is set at
$\Delta t=1$. Hence, the timescale separation is $\left(\nu\Delta
t\right)^{-1} \sim \nu^{-1}$. For slow driving (small $\nu$), SOC
theory expects the appearance of power laws. On the other hand, for
fast driving, a different trend is expected. We found this fit to
resemble a normal distribution, and is centered at a large $M$
(\textcolor{blue}{\tiny $\blacksquare$}, Fig.~\ref{fig:2}). An
implication of the crossover is that under high-rate driving
(especially by frequent rainfall), the landslide behavior of
mountain slopes appears to produce relatively large avalanches on
average. It is thus not surprising that highly-devastating
rain-induced landslides occur more often in areas frequently struck
by storms. However, a more comprehensive description of $\nu$ should
incorporate ground-failure factors such as soil composition. Such
factors have been neglected in this study for simplicity.

The rollover observed in our data supports the claim that the
rollover seen in substantially complete landslide-inventory datasets
is not a mere artifact of limited mapping resolution
~\citep{GuzzettiMalamudTurcotteReichenbach2002,MalamudTurcotteGuzzettiReichenbach2004a,MalamudTurcotteGuzzettiReichenbach2004b}.
As $\nu\rightarrow 0$ , the rollover is expected to disappear thus
leaving the power-law tail of the distribution. This agrees with
predictions from the Olami-Feder-Christensen model~\citep{OFC1992},
which is the limiting case of our model for $\nu\rightarrow 0$. To
the extent that our experiment and numerical results agree, we thus
attribute the rollover to the finiteness, albeit small, value of the
driving rate $\nu$ assumed to characterize real physical systems.
From this argument it can thus be reasonably deduced that the
rollover increases in prominence, such that the peak shifts towards
larger avalanche sizes, as $\nu$ increases. We found that at $\nu
\sim 7.5\times 10^{-4}$, the avalanche size distribution more
closely resembles a normal (or Gaussian) distribution that is
centered at a large size value.

\begin{figure}
\noindent\includegraphics[width=20pc]{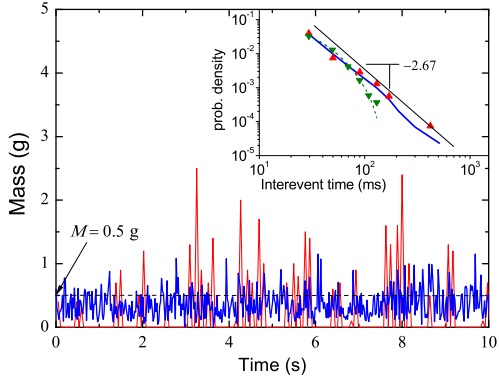}

\caption{(Color online). Interevent occurrence times (IOT). Time
series of avalanche sizes from experiment (red, solid curve) and
model (blue, thick curve). Consecutive data points are separated by
$20$~ms. Iteration steps (iter) in the model have been rescaled to
time by multiplying by the scaling factor: $0.05
\,\mathrm{s}\cdot\mathrm{iter}^{-1}$. Avalanche sizes less than
$0.5$~g are discarded. IOT is the interval between consecutive peaks
above the threshold. \emph{Inset}: IOT distributions for experiment
(\textcolor{red}{$\blacktriangle$}) and model (blue solid curve)
both exhibit a power-law with exponent $-2.670\pm 0.001$. Randomly
shuffling the order of the time series effectively results in an
exponential IOT distribution
(\textcolor[rgb]{0,0.5,0}{$\blacktriangledown$}; curve fit: green,
dashed curve).}\label{fig:3}
\end{figure}

A relevant aspect for hazard prediction is the interevent occurrence
time (IOT) statistics. The IOT is the interval between the peaks of
events whose sizes are above a given threshold ($M=0.5$~g), which
corresponds roughly to the peak of the pdf
(\textcolor{red}{\scriptsize $\blacklozenge$}, Figure~\ref{fig:2}).
We thus gathered time series of avalanche size $M(t)$ in
slowly-driven mounds over a period of 20 s. Figure~\ref{fig:3}
illustrates a representative sample for the first 10 s of this time
series for both experiment and computational data. A thresholding
procedure is applied to discard events whose sizes are below the
threshold. The region $M < 0.5$~g corresponds to the rollover
portion of the pdf (\textcolor{red}{\scriptsize $\blacklozenge$},
Figure~\ref{fig:2}) for slowly-driven mounds. The vertical
demarcation line in Figure~\ref{fig:2} corresponds to the horizontal
demarcation line in Figure~\ref{fig:3}. The probability density for
interevent occurrence time derived from experiment and model are
shown on the inset graph of Figure~\ref{fig:3}
(\textcolor{red}{$\blacktriangle$}; blue, solid curve), and both
exhibit a power-law trend with exponent $-2.670$($\pm 0.001$;
standard error of least-squares fit, $df=6$, reduced $\chi^2 =
2.09\times 10^{-8}$).

A power-law IOT distribution implies that most correlated events
(i.e., those with sizes above the threshold) tend to occur close
together in time, which seems to agree with recent
findings~\citep{Rossi2008}. The power-law trend further implies that
correlated events may be separated a long time from each other---a
landslide today may be linked with an earlier landslide a long time
ago. This may be tied to the fact that landslides tend to occur
where they have occurred before. The long temporal correlation
suggested by the power-law could possibly be attributed to rainfall
seasonal trends~\citep{Rossi2008}.

To illustrate that a power-law emerges from temporal correlations in
the time series data, a shuffling procedure has been implemented
wherein the order of the time series is rearranged
randomly~\citep{Yang2004}. Random shuffling effectively eliminates
any trace of correlations present in the original time series data.
The thresholding procedure is then applied on the shuffled data to
extract the IOT distribution. After shuffling, the IOT distribution
becomes exponential, as shown in Figure~\ref{fig:3}
(\textcolor[rgb]{0,0.5,0}{$\blacktriangledown$}, Inset). The change
from power-law to exponential due to shuffling has been expected for
systems governed by SOC dynamics~\citep{Woodward2004}. Therefore, we
conclude that the power-law is a direct indication of temporal
correlations in the time series data.

\emph{Conclusion}.---The general agreement of our results with
empirical data suggests that miniature experimental models may help
in understanding several underlying facets of complex landslide
processes. While several geophysical factors such as sand porosity,
rock type, pore-water pressure, and humidity have been neglected,
the landslide model presented here delivers basic insights that
would guide more detailed explorations later on.

\begin{acknowledgments}

The authors gratefully acknowledge B.D. Malamud for his insightful
comments on our manuscript. We also thank B. Buenaobra for
assistance on instrumentation; O. Burgos and M. Abundo for the
earthquake simulator; and funding from the UP-OVPAA (C.M.), UP-OVCRD
(D.E.J. and C.M.), and DOST (A.L. and R.B.).

\end{acknowledgments}

\end{article}

\end{document}